# Characterization of intra device mutual thermal coupling in multi finger SiGe:C HBTs

Mario Weiß[1], Amit Kumar Sahoo[1], Cristian Raya[2], Marco Santorelli[1], Sébastien Fregonese[1], Cristell Maneux[1] and Thomas Zimmer[1]

[1] Laboratoire IMS, CNRS - UMR 5218, Université de Bordeaux 1
[2] XMOD Technologies, Bordeaux, France
Email: mario.weisz@ims-bordeaux.fr

*Abstract*— This paper studies the mutual coupling in trench isolated multi-emitter bipolar transistors fabricated in a Si/SiGe:C HBT technology STMicroelectronics featuring $f_T$ and $f_{max}$ of ~300GHz and ~400GHz, respectively. Thermal coupling parameters are extracted using three-dimensional (3D) thermal TCAD simulations. The obtained parameters are implemented in a distributed transistor model that considers self-heating as well as thermal coupling between emitter fingers. Very good agreement is achieved between circuit simulations and DC measurements carried out on an in-house designed test structure.

## I. Introduction

Thermal issues in new high-speed submicron process technologies exacerbate the difficulties of performance prediction and reliability of next generation nonlinear RF circuits. Optimized device structure along with aggressive scaling leads to higher current densities and internal electric fields that increase the device and also the circuit operating temperature. In addition to their own self-heating, devices in close proximity can experience a temperature increase due to thermal coupling. Recent studies have demonstrated that self-heating and thermal coupling can have significant impact on circuit performance [1]–[3].

A common case where thermal coupling should be considered is in multi-transistor and multi-finger devices. The individual transistors/fingers, depending on trench isolation and proximity, can be strongly thermally coupled through the substrate. Multi-finger transistors are often used since they are more suitable for high frequency operation due to small base access resistance and capacitance. In this paper, intra device thermal coupling in multi-finger devices is studied using measurements on a special test structure as well as physical simulations. Thermal coupling parameters are extracted and implemented in a distributed transistor model.

## II. Test Structure

In order to accurately model the electro-thermal behavior of high power devices, test structures are required that help to understand self-heating, thermal coupling as well as transient thermal heating effects. In this section a multi-finger SiGe:C HBT structure is proposed for characterizing the thermal coupling between the emitter stripes. Each emitter has a drawn emitter area $A_E=[5\times0.18]\mu m^2$. The fingers are electrically connected as in the schematic diagram (Fig. 1a). In this configuration each finger transistor can be used either for heating or sensing. At the emitter node, the fingers are biased by independent voltage sources, controlling the power they dissipate. The common collector is pulled out which allows the base collector voltage to be significantly reverse biased in order to generate the desired heat. On the other hand, the $I_E(V_{BE})$ characteristic must be used for sensing which is less ideal than the $I_B(V_{BE})$. Each transistor is calibrated by measuring the I(V) characteristic at low current densities (negligible heat generation) at different chuck temperatures (27°C-110°C). Thermal interactions can be easily studied due to the various possible heat/sense combinations. The test structure is fully embedded in GSG pads to avoid oscillations as shown in the layout (Fig 1b).

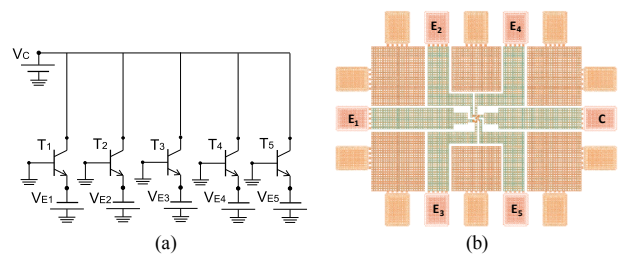

Fig. 1. Schematic diagram (a) and layout (b) of multi-emitter test structure.

## III. Electro-Thermal Simulations

The 3D nature of the heat flow inside transistors makes simple approximations of compact models of limited value. Predictive methods, therefore, require the accurate representation derived from numerical simulation. Thermal TCAD simulations are made with the Sentaurus device simulator using calibrated material properties, e.g. temperature dependent thermal conductivity. Due to the symmetry of the device, ½ of the five finger HBT structure is built on a semi-infinite Si-block with same geometrical dimensions as B5T technology. In the simulation, a power of 30mW is applied at five rectangular heat sources (placed at the base-collector junction of each finger). Fig. 2a shows the lattice temperature $T_{Lattice}$ distribution inside the device. A cross section of $T_{Lattice}$ along the heat source in x direction is displayed in Fig. 2b where a temperature difference between center and side fingers of $\Delta T=15K$ is observed.

## IV. Distributed Transistor Model

A single RC network is typically used to calculate the device temperature, although using higher order networks would lead to greater accuracy [4] in transient prediction. Netlist-based methods that calculate the thermal coupling between devices in an electrical simulator are most suitable for integrated circuit design. In this work, the thermal nodes of five single transistor models (HiCuM [5]) are connected to a distributed thermal network proposed by [6]. The thermal coupling terms were implemented by using voltage controlled voltage sources.

Each transistor model represents one of the five emitter fingers. The transistor model parameters are extracted from an HBT with CBEBC configuration presented in [3]. In order to adapt this model for each finger the external collector resistance $r_{cx}$ and the thermal resistance $R_{th}$ have been recalculated. The implemented thermal coupling parameters shown in Table I are extracted from physical simulation (see Section III).

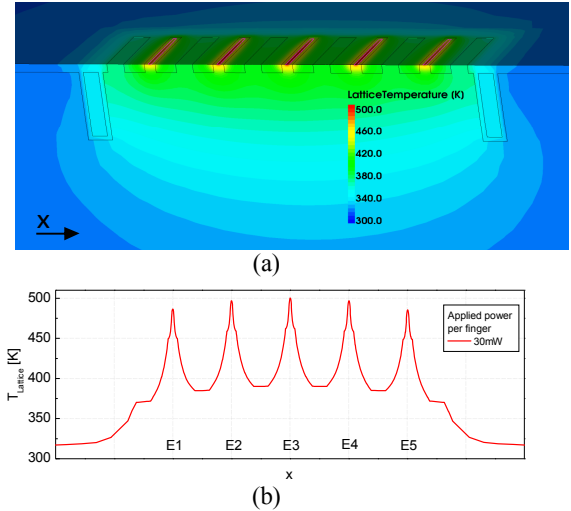

(a)

(b)

Fig. 2. (a) Lattice temperature $T_{Lattice}$ distribution inside the device when a power of 30mW is applied at five rectangular heat sources. (b) A cross section of $T_{Lattice}$ along the heat source in x direction.

TABLE I
COUPLING FACTORS

| Tran. Finger | Thermal coupling on | | | | |
|---|---|---|---|---|---|
| | T1 | T2 | T3 | T4 | T5 |
| T1 | 100% | 13% | 7.2% | 4.7% | 3.5% |
| T2 | 13% | 100% | 12.5% | 7.1% | 4.7% |
| T3 | 7.2% | 12.5% | 100% | 12.5% | 7.2% |
| T4 | 4.7% | 7.1% | 12.5% | 100% | 13% |
| T5 | 3.5% | 4.7% | 7.2% | 13% | 100% |

## V. RESULTS

On wafer measurements of the multi-finger test structure were carried out at room temperature. In Fig. 3, measured output characteristics for a base emitter voltage $V_{BE}$=[0.85, 0.9, 0.95]V are compared to circuit simulations using the extracted distributed thermal network. In Fig. 3a, emitter 1, 2 and 3 operate solely, which means that the differences are mainly caused by the difference in thermal resistance of each finger. The parallel operation of 2, 3 and 5 fingers is displayed in Fig 3b, 3c and 3d, respectively. Since the device structure is symmetrical, other combinations will not provide significant information. Very good accuracy is achieved with circuit simulations using the transistor compact models connected with the proposed thermal network.

## VI. CONCLUSIONS

HBTs have superior RF performance in comparison to standard CMOS, but circuit designers need accurate models to exploit their full potential. Therefore, a method which determines inter- and intra-device coupling has been applied to a state of the art SiGe:C multi-emitter HBT. Very good agreement between simulation and measurements has been achieved which makes this method ideally suited to electro-thermal simulation of high integrated circuits.

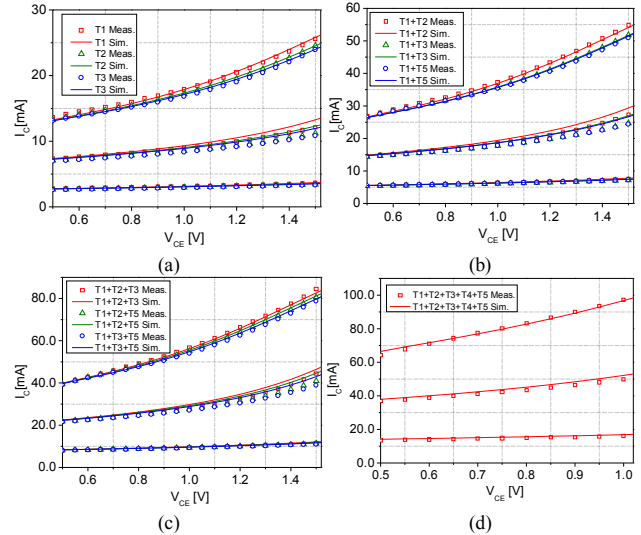

Fig. 3. Output characteristics of a multi-emitter HBT with $A_E$=0.27x5µm². Measurements and circuit simulations using a distributed network were carried out for a $V_{BE}$=[0.85, 0.9, 0.95]V and different finger combinations: (a) 1 emitter (b) 2 emitters (c) 3 emitters (d) 5 emitters operate in parallel.

ACKNOWLEDGMENT

This work is part of the RF2THZ SiSoC project supported by the European EUREKA Program CATRENE and of the DOTSEVEN project supported by the European Commission through the Seventh Framework Program for Research and Technological Development.